\pdfoutput=1

\documentclass[11pt]{article}

\usepackage[final]{acl}

\usepackage{times}
\usepackage{latexsym}

\usepackage{microtype}

\usepackage{inconsolata}
\usepackage[utf8]{inputenc} 
\usepackage[T1]{fontenc}    
\usepackage{hyperref}       
\usepackage{url}            
\usepackage{booktabs}       
\usepackage{amsfonts}       
\usepackage{nicefrac}       
\usepackage{microtype}      
\usepackage{xcolor}         
\usepackage{xnewcommand}
\usepackage{enumitem}
\usepackage{hyperref}
\hypersetup{
    colorlinks=true,
    citecolor=blue,
    linkcolor=blue,
    urlcolor=blue,
}

\usepackage{amsmath}
\usepackage{cleveref}
\usepackage{wrapfig}

\usepackage{listings}
\usepackage{xcolor}
\usepackage{graphicx}
\usepackage{multirow}
\usepackage{diagbox}
\usepackage{amssymb}
\usepackage{longtable}
\usepackage{subcaption}

\definecolor{codegreen}{rgb}{0,0.6,0}
\definecolor{codegray}{rgb}{0.5,0.5,0.5}
\definecolor{codepurple}{rgb}{0.58,0,0.82}
\definecolor{backcolour}{rgb}{0.95,0.95,0.92}

\lstdefinestyle{mystyle}{
    backgroundcolor=\color{backcolour},   
    commentstyle=\color{codegreen},
    keywordstyle=\color{blue},
    numberstyle=\tiny\color{codegray},
    stringstyle=\color{codepurple},
    basicstyle=\ttfamily\footnotesize,
    breakatwhitespace=false,         
    breaklines=true,                 
    captionpos=b,                    
    keepspaces=true,                 
    numbers=left,                    
    numbersep=5pt,                  
    showspaces=false,                
    showstringspaces=false,
    showtabs=false,                  
    tabsize=2
}

\title{$\pname$: A Benchmark to Evaluate Large-Language Models for Assertion Generation for Hardware 
Design}

\author{
  \textbf{Vaishnavi Pulavarthi\textsuperscript{1}},
  \textbf{Deeksha Nandal\textsuperscript{1}},
  \textbf{Soham Dan\textsuperscript{2}\thanks{SD and DP jointly supervised this work.}},
  \textbf{Debjit Pal\textsuperscript{1}\footnotemark[1]}
\\
\\
  \textsuperscript{1}Dept. of Electrical and Computer Engineering, \\
  University of Illinois Chicago, Chicago IL 60607,\\
  \textsuperscript{2}Microsoft,
\\
  \small{
    \textbf{Correspondence:} \href{mailto:dpal2@uic.edu}{dpal2@uic.edu}
  }
}

\newcommand{\pname}{\mbox{{AssertionBench}}}

\newcommand{\goldmine}{\mbox{{\scshape GoldMine}}}
\newcommand{\gptt}{\mbox{{GPT-3.5}}}
\newcommand{\gptf}{\mbox{GPT-4o}}
\newcommand{\cllama}{\mbox{CodeLLaMa 2}}
\newcommand{\llama}{\mbox{LLaMa3-70B}}

\newcommand{\bem}[1]{{\bf\em #1}}

\newcommand{\eg}{\mbox{{\em e.g.}}}
\newcommand{\ie}{\mbox{{\em i.e.}}}
\newcommand{\cf}{\mbox{{c.f.}}}

\newcommand{\proc}[1]{\ifmmode\mbox{\textsc{#1}}\else\textsc{#1}\fi}

\renewcommand{\paragraph}[1]
{\noindent \textbf{#1} --%
}

\newlength{\Oldarrayrulewidth}

\definecolor{mygreen}{HTML}{e0f3db}
\definecolor{myblue}{HTML}{a8ddb5}
\definecolor{myorange}{HTML}{43a2ca}

\begin{document}

\maketitle


\begin{abstract}
    Assertions have been the de facto collateral for hardware verification 
    for over a decade. The verification quality, 
    \ie, detection and diagnosis of corner-case design bugs, is critically dependent on the assertion quality. 
    There has been a considerable amount of research 
    to generate high-quality assertions from hardware design source code and design execution trace data. With the recent advent of generative AI techniques such as Large Language Models (LLMs), there has been a renewed interest in deploying LLMs for assertion generation. 
    However, there is little effort to quantitatively establish the effectiveness and suitability of various LLMs for assertion generation. In this paper, we present $\pname$, a novel benchmark to evaluate LLMs' effectiveness for assertion generation quantitatively. $\pname$ contains $100$ curated Verilog hardware designs from OpenCores and formally verified assertions for each design, generated from $\goldmine$ and HARM. We use $\pname$ to compare state-of-the-art LLMs to assess their effectiveness in inferring functionally correct assertions for hardware designs. Our experiments comprehensively demonstrate how LLMs perform relative to each other, the benefits of using more in-context examples in generating a higher fraction of functionally correct assertions, and the significant room for improvement for LLM-based assertion generators. 
\end{abstract}

\section{Introduction: 
}\label{sec:intro}

System-on-Chip (SoC) designs are 
the building blocks for many safety-critical computing systems, 
including vehicular systems, infrastructure, military, and industrial automation. 
It is crucial 
to establish 
that the SoCs are functionally correct, safe, and secure to ensure that the SoC designs function as intended and are free from errors and vulnerabilities. 

\bem{Assertions} or {\em design invariants} are mathematical encoding (in Boolean logic) of desired design properties that should hold True 
for the hardware design. 
Assertion-based Verification (ABV)~\cite{witharana2022assertion} has long emerged as the de facto standard to verify the security and functional correctness 
of SoCs. However, crafting a succinct set of assertions for hardware designs is a tedious and time-consuming task, often requiring considerable human ingenuity. Too many assertions can negatively affect 
verification performance with a prolonged verification closure, whereas too few assertions may result in insufficient design coverage causing corner case design bugs to escape to production and mass manufacturing. 
\bem{Consequently, it is crucial to develop 
automated and scalable techniques to rapidly generate a succinct set of hardware design properties targeting design functionality, safety, and security}.

There has been a considerable amount of research work for assertion generation using 
two different paradigms -- lightweight static analysis of design source code and formal verification 
~\cite{
tiwari2004tacas, spin2004}, 
and data-driven statistical analysis, \eg, data mining~\cite{daikon2000ernst, iodine2005, pinter2005hase, inferno2009tcad, chang2015aspdac, wang1998dac, hekmatpour2005, rogin2008date, inferno2009tcad, chang2015aspdac, chung2011iccd}. 
$\goldmine$, for the first time developed a static analysis guided statistical analysis-based technique to generate hardware assertions~\cite{goldmine, hertz2013tacd} in Linear Temporal Logic~\cite{ltl} in a fully automated way. While $\goldmine$ and follow-up research~\cite{hertz2019aspdac, pal2020tcad, malburg2017aspdac, iman2024artmine, iman2021norcas, danese2017dac, germiniani2022vlsi, iman2022dsd, harm2022tcad, calvin2021ashes, hasini2023jetc, avinash2024host, pal2020tcad, iman2022dsd, tara2015vlsisoc} 
made assertions accessible beyond the hardware verification engineers, most of those methods still failed to scale to large industry-scale designs due to the algorithmic complexity of the underlying static analysis.

With recent advances in 
generative AI models, especially Large Language Models (LLMs), 
there is a renewed interest in the research community to harness the power of LLMs for assertion generation task. Most recent assertion generation approaches~\cite{liu2023verilogeval, orenesvera2023using, kande2023llmassisted, fang2024assertllm, mali2024chiraag, stutton2023icml} treat LLMs as \bem{black-box} and use {\em prompt engineering} to iteratively refine the set of generated assertions. 
However, there is no in-depth study nor a dataset to evaluate how well different state-of-the-art (SOTA) LLMs perform on generating the correct set of assertions without 
designer developed prompts.

To address this gap, we propose $\pname$: the first comprehensive benchmark to quantify the efficacy of SOTA LLMs for the assertion generation task. 
The benchmark contains $100$ curated designs of varying complexity encompassing a broad spectrum of design types, including 
encoders, decoders, and arithmetic operations such as addition, multiplication and 2's complement in 
Floating Point Units, 
along with their formally verified assertions facilitating future research in exploring the applicability of LLMs in assertion generation. In this work we quantify the quality 
of the LLM-generated assertions prompted with a collection of labeled designs (and their formally verified assertions) as in-context learning (ICL) examples.


\section{The $\pname$ Benchmark 
}\label{sec:benchmark}



%

$\pname$\footnote{\url{https://github.com/achieve-lab/assertion_
data_for_LLM.}} contains 
ICL example designs and 
and test designs from OpenCores~\cite{opencores}. 
The benchmark contains 
combinational and sequential hardware designs 
containing up to 1250 lines of codes (LoCs)~\cite{cloc} excluding comments and blank lines. 

In our benchmark, each ICL 
example 
consists of a Verilog design and its formally verified assertions, generated from $\goldmine$~\cite{pal2020tcad} and HARM~\cite{harm2022tcad}, and verified using Cadence JasperGold~\cite{jaspergold}. 
$\pname$ primarily contains 
Verilog designs for benchmarking for the following reasons.   

\begin{enumerate}[label=(\alph*), leftmargin=*, parsep=0cm, itemindent=0em, itemsep=0.2em, topsep=0.1em]
    \item The most recent work on LLM-assisted hardware designs focuses on Verilog, the predominant language for hardware design in industry and academia. For example, recent works such as VerilogEval \cite{liu2023verilogeval}, MG-Verilog \cite{mgverilog}, Isadora~\cite{calvin2021ashes} and HARM \cite{harm2022tcad} 
    solely focus on Verilog. Our work 
    aligns with the predominant and widely used hardware design language (HDL) in the state-of-the-art research and practice. 

    \item To our knowledge, no public domain assertion generation tool is available to generate assertions for VHDL/SystemC designs. The only work that mines assertions from SystemC that we can find is~\citet{liu2011memocode}. However, we could not find the actual implementation of the tool in the public domain. Such assertions are necessary as ICL examples. This emphasizes the value of $\pname$ in complementing existing research and in future work we plan to augment it with VHDL/SystemC designs and corresponding assertions.
\end{enumerate}

The key research question (RQ) that $\pname$ aims to address is \bem{whether we can effectively leverage state-of-the-art (SOTA) deep learning (DL) tools}, such as LLMs, to assist verification engineers in crafting assertions for large hardware designs and overcome the shortcomings of the classical static and dynamic analysis based techniques. There are ongoing efforts to leverage LLMs to alleviate the shortcomings. Our effort in designing $\pname$ is the first step towards ensuring that as we develop LLM-assisted techniques for assertion generation, we remain aware of the insights garnered by this work and avoid pitfalls. This benchmark and the framework will allow us to quantitatively and systematically compare the fitness of existing and future closed and open-sourced LLMs for the assertion generation task.  

The ICL example 
set comprises fundamental sequential designs (aka designs with a clock) such as Arbiter, T Flip-Flop, and combinational designs (aka designs without a clock) such as Half Adder, Full Adder, and Full Subtractor. 
We consider the 
corresponding concurrent assertions (sequence subset of SVA) for ICL, and the 
which contain 
overlapped and non-overlapped implication operators~\cite{svmlrm}. Our test design set consists of highly curated 100 Verilog designs (both sequential and combinational designs) of varying complexity 
that are up to 20$\times$ bigger 
than those in the ICL examples 
in term of LoCs, 
to evaluate 
LLMs' 1-shot and 5-shot assertion generation ability. We wanted to understand if LLMs can learn about an assertion and how it relates an assertion to the design source code using simple examples. \bem{Our main goal was to evaluate if the LLMs can transfer the learned knowledge successfully to much bigger designs}. A successful demonstration of such transfer of the learned knowledge would pave the path that LLMs can be an effective tool for assertion generation at scale, likely for industrial-scale designs even when learned and/or trained using smaller design source codes. We also wanted to understand \bem{if LLMs would excel or struggle for the restricted subset of assertion classes}. Our results (\cf,~\Cref{sec:exp_res}) show that LLMs struggle to generate correct assertions for this restricted subset. Hence it is futile to delve further into more complex assertion constructs 
unless we have clear insight into why LLMs are failing for the restricted subcases. In this effort, we develop those insights.

\section{Experimental Setup 
}\label{sec:exp_setup}

\begin{figure}
    \centering
    \includegraphics[width=\columnwidth]{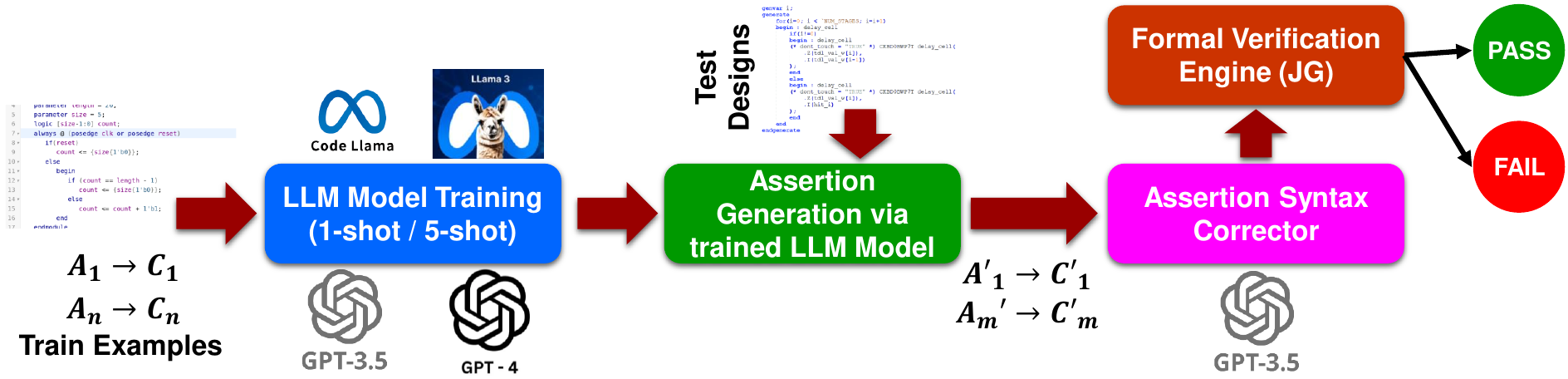}
    \vspace{-4mm}
    \caption{{\bf Our evaluation framework}. {\bf JG}: 
    JasperGold Formal Property Verification Engine.}
    \label{fig:eval_framework}
    \vspace{-5mm}
\end{figure}

\Cref{fig:eval_framework} shows our evaluation framework. We evaluated four SOTA LLMs 
$\gptt$~\cite{chatgpt35}, $\gptf$~\cite{openai2024gpt4}, $\cllama$~\cite{codellama2}, and $\llama$~\cite{llama3} using the proposed $\pname$ on the task of predicting correct or valid assertions.

\smallskip

\noindent {\bf Compute Platform}: 
We use UIUC 
NCSA's 
public Delta Cluster~\cite{ncsadelta} 
for our experiments. 
We use GPU nodes containing 1-way, 4-way, and 8-way NVIDIA A40 (with 48GB GDDR6) and A100 (with 40GB SXM) GPUs to perform $k$-shot learning 
and inference. Each 1-way and 4-way GPU computing node has 256 GB RAM, and each 8-way GPU computing node has 2 TB RAM. 

\smallskip

\noindent {\bf Hyperparameters}: We use default hyperparameters for all the LLM models under consideration. 
Specifically, the number of maximum output tokens has 
been set 
at 1024, employing a greedy decoding strategy and maintaining a \textit{temperature} of 1.0, \textit{top\_p} of 0.95. The \textit{random seed} has been configured to 50.

\begin{figure}
\begin{lstlisting}[language=Verilog, 
                 linewidth=\columnwidth, framexleftmargin=2pt,
                 framexrightmargin=2pt,
                 ]
You are an expert in SystemVerilog Assertions.
Your task is to generate the list of assertions to the given verilog design. An example is shown below. Generate only the list of assertions for the test program with no additional text.
Program 1: module arb2(clk, rst, req1, req2, gnt1, gnt2); input clk, rst; ...
Assertions 1: (state == 1 & req2 == 1) |-> (gnt1 == 0);... 
Test Program: 
module fifo_mem #(parameter DEPTH=8, DATA_WIDTH=8, PTR_WIDTH=3) ( input wclk, w_en, rclk, r_en, input [PTR_WIDTH:0] b_wptr,  ...
Test Assertions:
\end{lstlisting}
\vspace{-4mm}
\caption{{\bf An example of the prompt for 1-shot learning}. The example consists of a tuple, a Verilog 
design ({\tt Program 1}) and a set of formally verified assertions for the design ({\tt Assertions 1}). The {\tt Test Program} is the Verilog 
design for which we generate assertions using the trained LLM.}
\label{fig:prompt_1shot}
\vspace{-5mm}
\end{figure}


\smallskip 

\noindent {\bf Pre-trained Models and Verification 
Tool}: We use pre-trained LLMs from the HuggingFace model repository~\cite{huggingface} for 
$\pname$ evaluation. We have also used Python 3.11 and Cadence JasperGold (JG) version 2022.06p002 for formally verifying the assertions generated from the test Verilog designs. We use two SOTA classical tools $\goldmine$~\cite{pal2020tcad} and HARM~\cite{harm2022tcad} to generate {\em example assertions} from different Verilog designs in the IC set. Below, we summarize the four LLMs that we evaluate using 
$\pname$. 

\smallskip

\noindent $\bullet$ {\bem{$\gptt$}} is 
a commercial autoregressive LLM (from OpenAI) based on the GPT architecture, pre-trained on extensive text corpora and fine-tuned for NLP tasks.

\smallskip

\noindent $\bullet$ {\bem{$\gptf$}} is a unified multimodal transformer model processing text, vision, and audio, with enhanced reasoning and programming capabilities over previous iterations.

\smallskip

\noindent $\bullet$ {\bem{$\cllama$}} is 
a suite of autoregressive transformer models for code and text generation, including a 70B parameter variant that uses Grouped-Query Attention for scalable inference.

\smallskip

\noindent $\bullet$ {\bem{$\llama$}} is 
a 70B-parameter transformer with an 8,192-token context window, pre-trained on 15 trillion tokens from publicly available datasets.

\smallskip 

\noindent {\bf Evaluation Protocol}: To evaluate effectiveness of the different LLMs, 
the few-shot testing regime consists of 1-shot and 5-shot ICL 
examples. Each 
example is a tuple consisting of 
a Verilog design source code and 
a set of formally verified assertions containing up to 
10 assertions.  
~\Cref{fig:prompt_1shot} shows our prompt 
consisting of four parts -- (i) an English language description of the task, 
(ii) the Verilog design,  
(iii) an assertion in SystemVerilog Assertion (SVA)~\cite{svmlrm} format, 
and (iv) a test Verilog design. 
Next, we prompt each LLM with the ICL examples and evaluate them on 100 test Verilog designs 
to infer assertions. In our experiments, we have found all of the LLMs generate syntactically erroneous assertions, \ie, each LLM fails to learn the SVA syntax from the ICL examples. Consequently, we use a syntax corrector (SC) 
using $\gptt$ and feed the output of the SC 
to JG 
to evaluate the quality of the generated assertions. 
Any other SVA-compatible formal property verifier (FPV) 
will work as well.

\smallskip

\noindent {\bf Metrics}: We evaluate the assertions generated for the test programs using 
the following three metrics for each LLM -- (i) {\bf Pass} quantifies the fraction of generated assertions that FPV 
attests as valid for the design; (ii) {\bf Fail} quantifies the fraction of generated assertions that FPV 
attests as wrong with a counterexamples trace (CEX); and (iii) {\bf Error} quantifies the fraction of generated assertions for which the FPV 
identifies syntactic errors 
even after syntax correction. We have not reported any other metric, \eg, assertion coverage~\cite{athavale2014dac}, as that is meaningful only when an assertion is valid 
and one wants to quantify the quality of the 
assertion or would like to induce a ranking on assertions~\cite{pal2020tcad}. 
In current work, we did not target to quantify the quality of the 
assertions neither did we want to induce a rank on them. 
Rather we focused on the ability of the SOTA LLMs on generating correct assertions. 

\section{Experimental Results 
}\label{sec:exp_res}

\begin{figure*}
    \centering
    \begin{subfigure}[b]{0.3\textwidth}
        \includegraphics[scale=0.2]{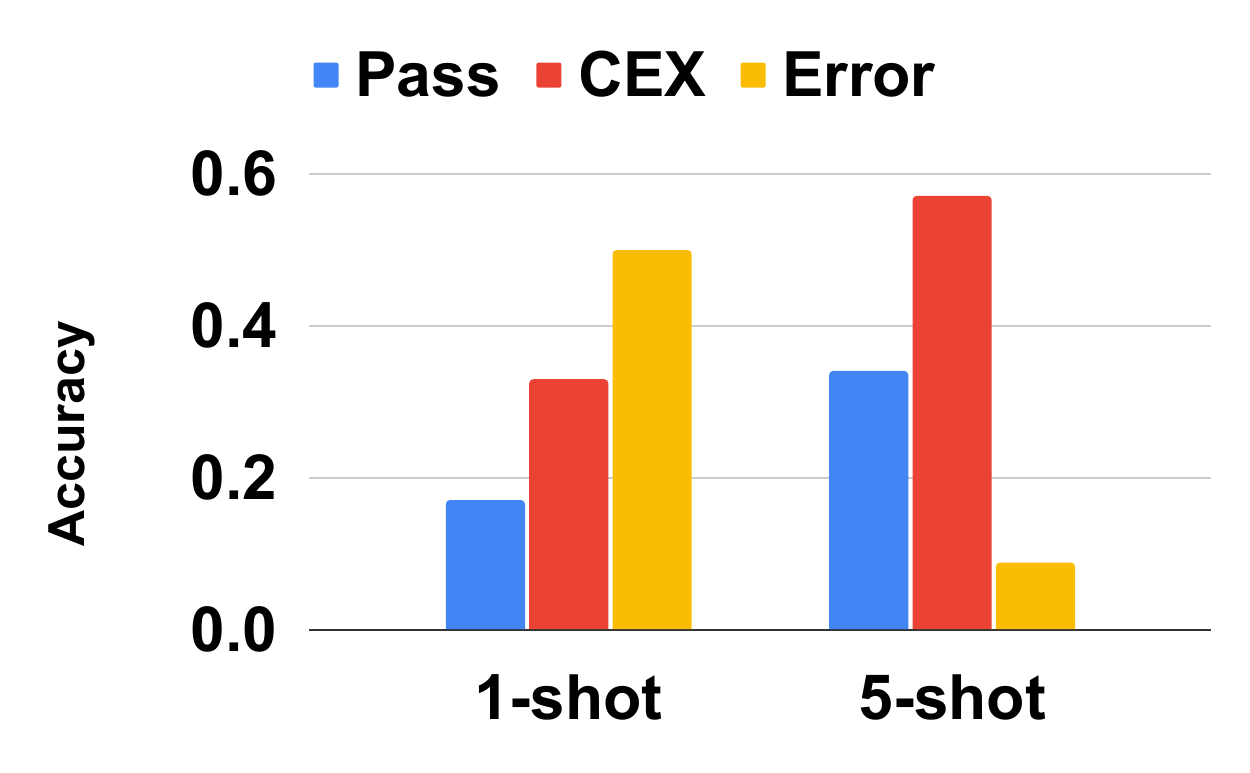}
        \vspace{-2mm}
        \caption{
        \label{fig:gpt35}}
    \end{subfigure}
    \hfill
    \begin{subfigure}[b]{0.3\textwidth}
        \includegraphics[scale=0.2]{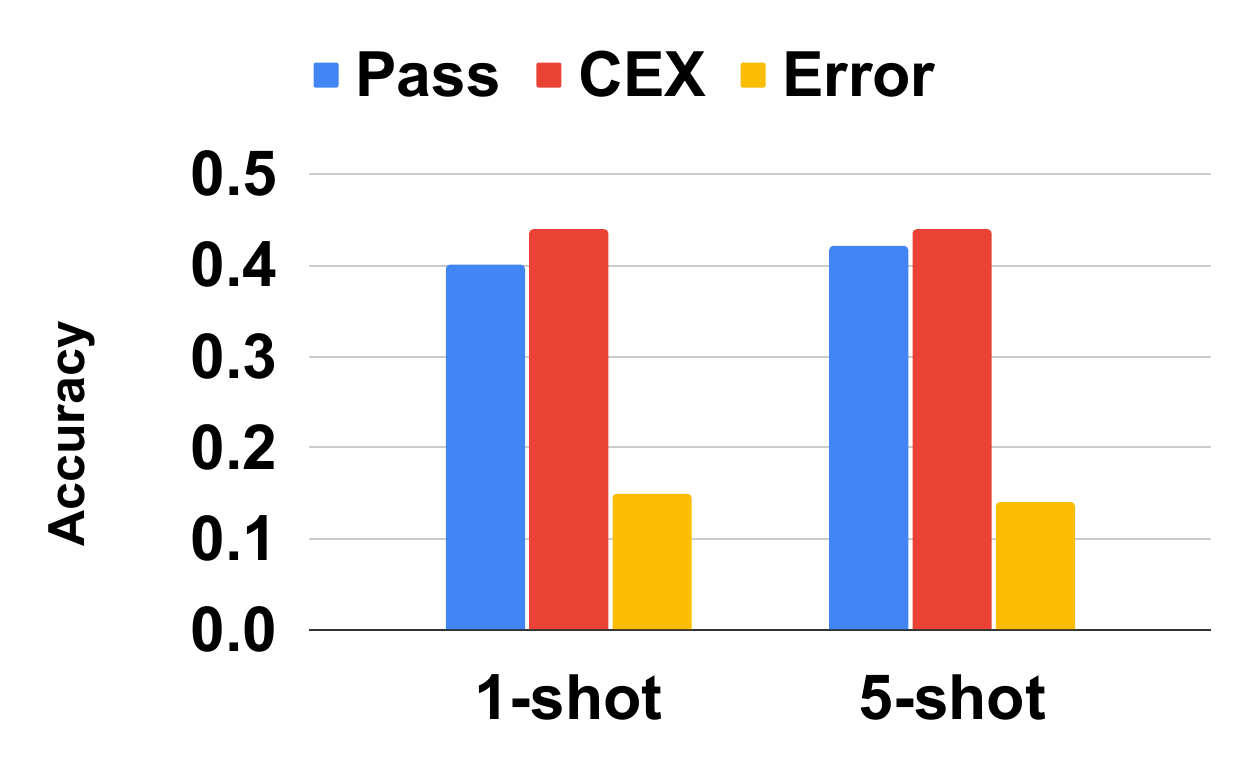}
        \vspace{-2mm}
        \caption{
        \label{fig:gpt4o}}
    \end{subfigure}
    \hfill
    \begin{subfigure}[b]{0.3\textwidth}
        \includegraphics[scale=0.2]{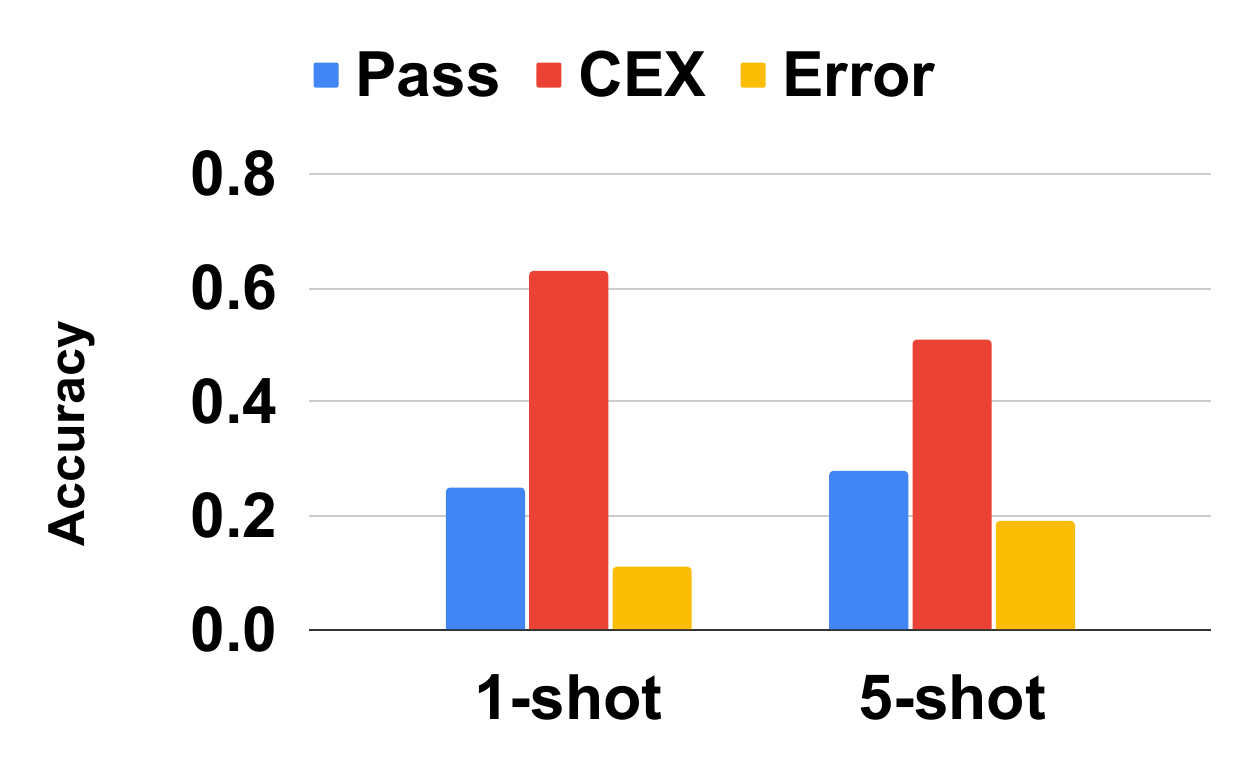}
        \vspace{-2mm}
        \caption{
        \label{fig:codellama2}}
    \end{subfigure}
    \begin{subfigure}[b]{0.3\textwidth}
        \includegraphics[scale=0.2]{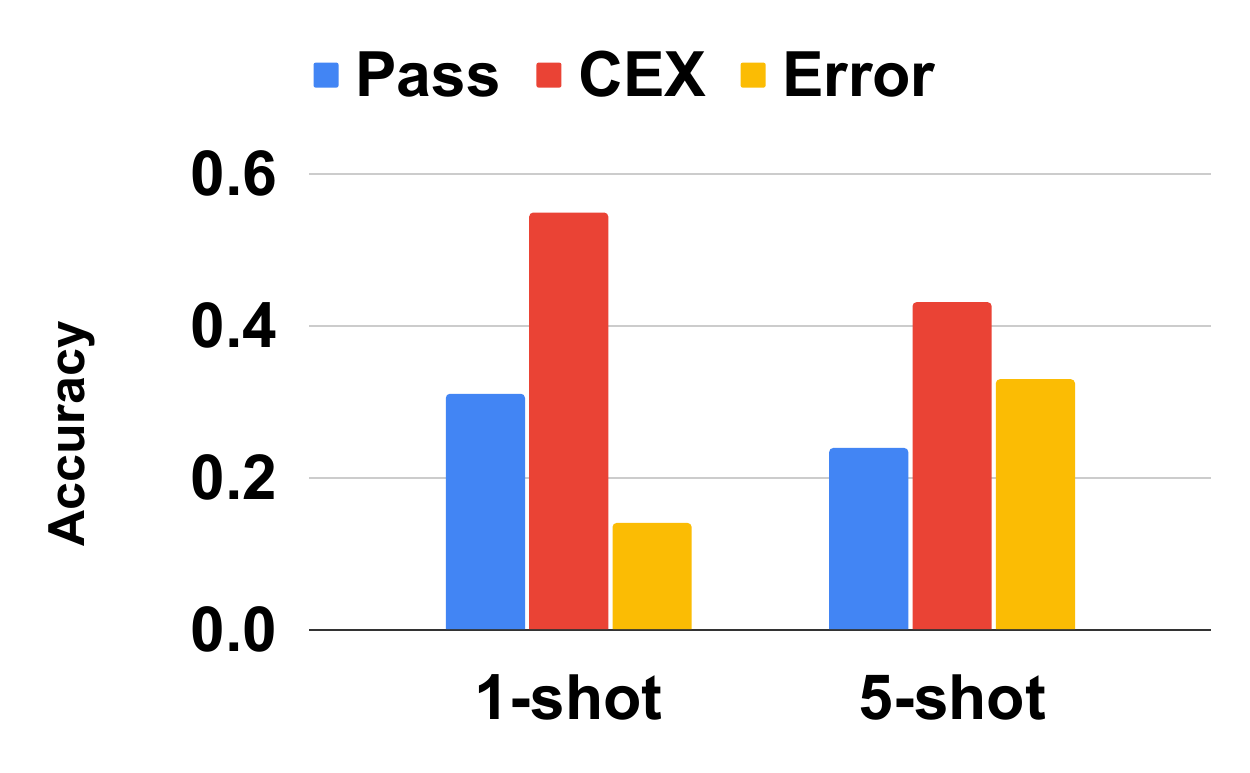}
        \vspace{-2mm}
        \caption{
        \label{fig:llama3}}
    \end{subfigure}
    \hfill
    \begin{subfigure}[b]{0.3\textwidth}
        \includegraphics[scale=0.18]{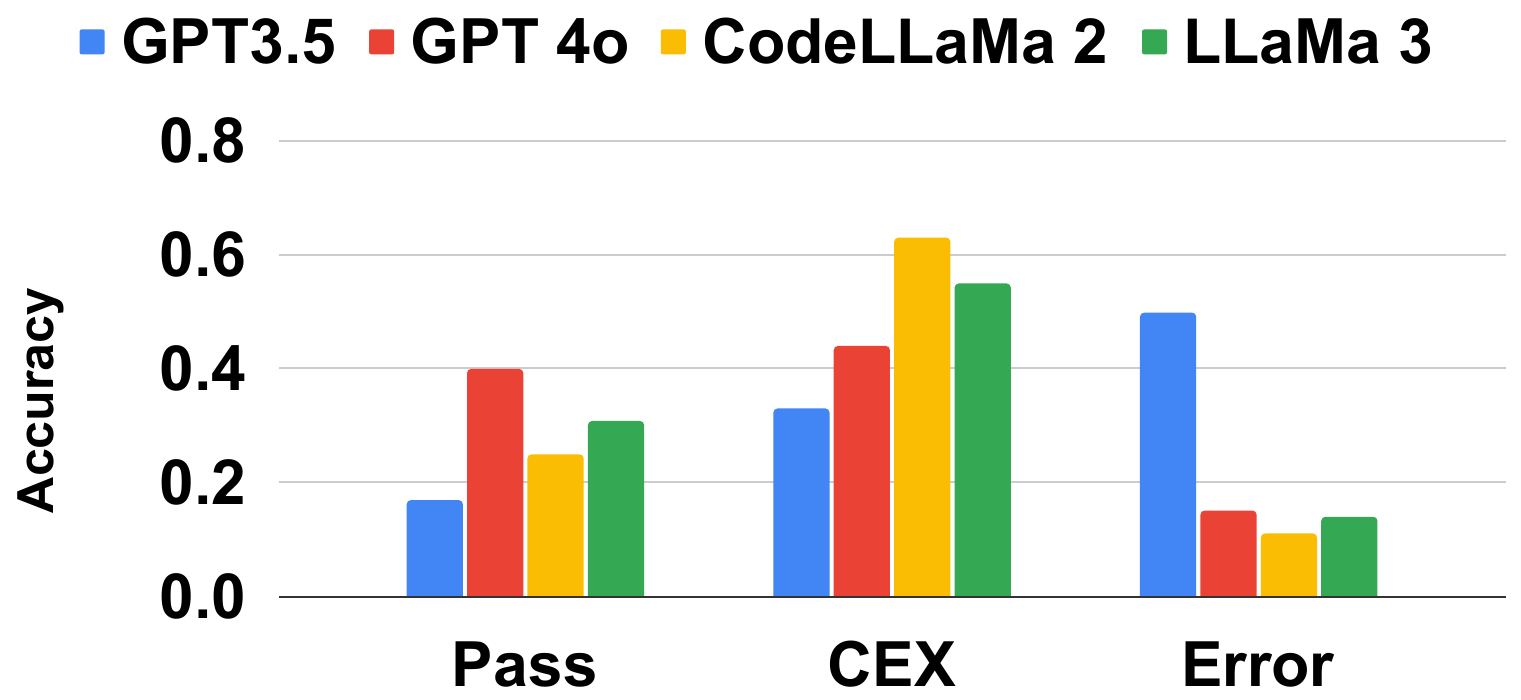}
        \vspace{-2mm}
        \caption{
        \label{fig:1_shot_acc}}
    \end{subfigure}  
    \hfill
    \begin{subfigure}[b]{0.3\textwidth}
        \includegraphics[scale=0.18]{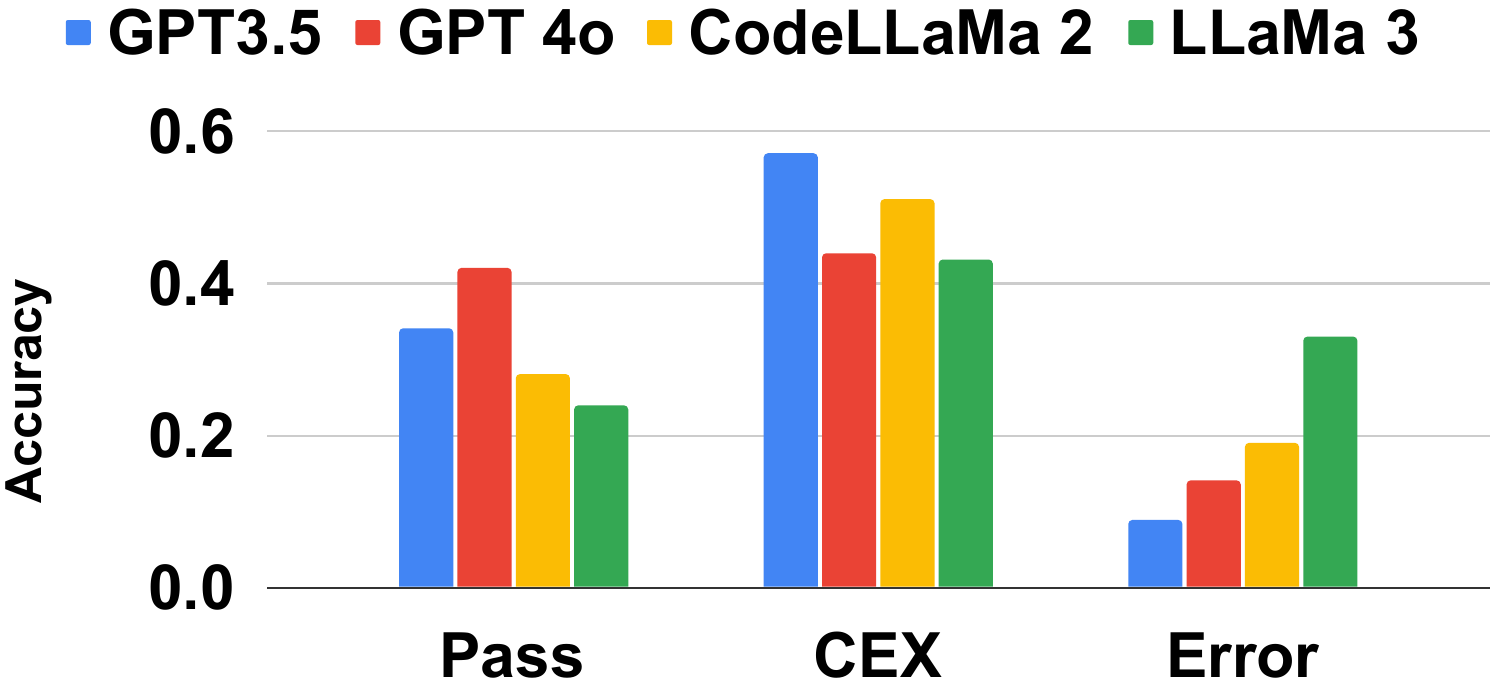}
        \vspace{-2mm}
        \caption{
        \label{fig:5_shot_acc}}
    \end{subfigure}
    \vspace{-3mm}
    \caption{{\bf Comparison of accuracy of generated assertions}. (\subref{fig:gpt35}) Assertion accuracy comparison for $\gptt$. (\subref{fig:gpt4o}) Assertion accuracy comparison for $\gptf$. (\subref{fig:codellama2}) Assertion accuracy comparison for $\cllama$. (\subref{fig:llama3}) Assertion accuracy comparison for $\llama$. (\subref{fig:1_shot_acc}) $k = 1$-shot assertion accuracy. (\subref{fig:5_shot_acc}) $k = 5$-shot assertion accuracy. {\bf CEX}: Counterexamples trace.
    }
    \label{fig:imp_between_shots}
    \vspace{-5mm}
\end{figure*}

We show our overall experimental results in~\Cref{fig:imp_between_shots}. 
We make following observations.

\smallskip

\noindent \bem{Observation 1}: {\bf Most LLMs generate valid assertions with an increasing number of ICL 
examples} (\cf,~\Cref{fig:gpt35}-\ref{fig:llama3}). 
$\gptt$, $\gptf$, and $\cllama$ show on average an improvement of 2$\times$, 1.2$\times$, and 1.12$\times$ for valid assertion generation, respectively, when moved from 1-shot learning to 5-shot learning. However, $\llama$ model loses accuracy from 31\% to 24\% on the same dataset. Our 
analysis shows in many cases, $\llama$ either fails to generate assertions or generates syntactically wrong assertions 
or tries to generate codes in a new programming language (\eg, 
Java). This experiment shows that {\em there is considerable scope for 
improving the $\llama$ model for this task, likely via fine-tuning the pre-trained $\llama$ model}. 

\smallskip

\noindent \bem{Observation 2}:  {\bf An enhanced model does not necessarily ensure a better semantic or syntactic understanding}.
For $\gptt$ (\cf,~\Cref{fig:gpt35}), with an increase in the number of ICL examples, the LLM was able to produce more syntactically correct assertions, however, 
majority of corrected assertions (on average up to 24\%) generated a CEX 
when verified with JG. 
For $\gptf$, the results were more consistent in terms of syntactically correct 
and failing assertions from 
1-shot and  5-shot learning (\cf,~\Cref{fig:gpt4o}). For $\cllama$ and $\llama$, with increase in the number of ICL examples, the fraction 
of failed assertions decreased (on average up to 12\% for $\cllama$ and $\llama$, \cf,~\Cref{fig:codellama2} and~\Cref{fig:llama3}), however, both models generated more syntactically wrong assertions (on average up to 19\% more for $\llama$). 
Our 
analysis shows that with 1-shot, the variation in types of assertions in the ICL examples were limited. Consequently, $\llama$ learned the syntax. However, in 5-shot learning, we have more variations in assertion syntax 
which made $\llama$'s learning task difficult. This experiment shows that {\em increasing the number of ICL examples will not necessarily improve LLM's consistency in generating passing, failing, and syntactically correct assertions}. Further, our analysis shows that {\em LLMs that are more performant on standard LLM benchmarks does not necessarily have a better semantic understanding when it comes to assertion generation}.

\smallskip

\noindent \bem{Observation 3}: {\bf $\gptf$ is relatively more consistent and superior for assertion generation task} (\cf,~\Cref{fig:1_shot_acc}-\ref{fig:5_shot_acc}). 
Our experiment shows that $\gptf$ generates 
on an average up to 15.6\% more valid assertions as compared to other LLMs for both 1-shot and 5-shot learning. Additionally, 
$\gptf$ produced fewer CEX 
generating assertions and syntactically incorrect assertions as compared to other LLMs. This experiment shows 
that $\gptf$ {\em is more beneficial for assertion generation 
as compared to the other LLMs}. 

\smallskip

\noindent \bem{Observation 4}: {\bf All LLMs need considerable improvement for assertion generation task} (\cf,~\Cref{fig:imp_between_shots}). Our 
analysis 
shows that none of the LLM models can generate valid assertions 
an average of no more than 44\% accuracy whereas up to 63\% generated assertions produce CEX 
and on average up to 33\% of generated assertions are syntactically wrong. Clearly, for LLMs to be of practical usage for any realistic industrial-scale design, considerable improvement needs to be made. Specifically, {\em the LLMs need to capture the semantic meaning of the Verilog designs for automatically producing a higher fraction of valid assertions 
without iterative human prompting}. 

\smallskip

\noindent \bem{Remarks}: In this work, we have refrained from reporting the coverage of assertions. We emphasize that unlike code-based coverage metrics, \eg, line / statement coverage, branch coverage, condition coverage, FSM coverage, etc., there is no well-defined notion of assertion coverage. To the best of our knowledge, the only work that connects assertion's coverage of the design code is by ~\citet{athavale2014dac}. They defined correctness-based coverage of an assertion as identifying the design statements / codes that contribute to its non-vacuous satisfaction. However, such coverage, \ie, assertion coverage as defined by ~\citet{athavale2014dac}, only makes sense when an assertion is correct (\ie, valid) and one wants to quantify the quality of the correct assertion or would like to induce a ranking on assertions, \eg,~\citet{pal2020tcad, tara2015vlsisoc}. In current work, we did not target to quantify the quality of the generated assertions and neither did we want to induce a rank on generated assertions; rather we focused on the fitness of the current and future commercial and open-source LLMs on generating correct assertions. Such coverage (and ranking) would make perfect sense if we pursue the overarching goal of developing task-specific LLMs for assertion generation to quantify the quality of the LLM-generated assertions which in turn would quantify the quality of the task-specific LLMs.

\section{Conclusion and Future Work 
}\label{sec:conclusion}


This work introduces $\pname$ to evaluate the current and future commercial and open-source LLMs for the assertion generation task. No prior work comprehensively benchmarks SOTA LLMs for assertion generation, especially for HDLs. To our knowledge, $\pname$ is the first such benchmark to quantitatively compare various LLMs in terms of goodness for the task of assertion generation. Although there is no LLM that consistently outperforms other LLMs, we notice several promising trends and research directions to enhance the practical applicability of LLMs for assertion generation task, which will further accelerate SoC and hardware design verification. As LLM research is growing at a tremendous pace both in commercial and academic research, we plan to maintain the benchmark and augment its learning and test set with more complex designs and their formally verified assertions to further stress test models. 
\clearpage
\section{Limitations}


We identify the following limitations of this work in terms of the dataset and the evaluation methodology.

\begin{itemize}
    \item {\bf Dataset}: In the scope of this study, our primary focus is on Verilog designs, given its status as the predominant hardware design language. Moving forward, it will be intriguing to develop benchmarks for assertions in alternative hardware languages, \eg, VHDL, SystemVerilog, and SystemC, thereby expanding the scope of our analysis to encompass a broader range of design paradigms. Additionally, $\pname$ considers only a few temporal assertions with shallow temporality. It would be interesting 
    to increase the temporal depth to capture design behaviors that cut across multiple clock cycles and evaluate LLM's ability to learn and generate assertions to capture such behaviors succinctly.
    
    \item {\bf Assertion Objective}: In this work, we primarily focused on the assertions that capture design functionality. It would be interesting to enhance and augment $\pname$ with security assertions to evaluate the LLM's ability to capture and summarize security violations/concerns from hardware design source code.
    
    \item {\bf Quantitative Assertion Ranking}: In this work, we primarily focused on correctness of an assertions without quantifying and ranking the subtlety of the captured design behavior~\cite{pal2020tcad}. It would be interesting to include such rankings in the ICL 
    examples and evaluating LLM's capability to automatically rank generated assertions to quantify captured design behavior. 
    
    \item {\bf Modeling}: In this paper, we assessed the few-shot assertion generation capabilities of SOTA 
    language models. In future work, it will be interesting to fine-tune language models for assertion generation and evaluate their performance on $\pname$.
    
    \item {\bf Evaluation}: In future work, it will be valuable to conduct a more detailed evaluation of model errors to better understand the specific limitations of each LLM for assertion generation.
\end{itemize}




\end{document}